\documentstyle[12pt]{article}

\newcommand{\be}{\begin{equation}}
\newcommand{\ee}{\end{equation}}
\newcommand{\bea}{\begin{eqnarray}}
\newcommand{\eea}{\end{eqnarray}}
\newcommand{\nono}{\nonumber}
\newcommand{\ci}{\cite}
\newcommand{\bi}{\bibitem}
\newcommand{\r}{{\bf r}}
\newcommand{\bfeta}{\mbox{\boldmath$\eta$}}
\newcommand{\bfnabla}{\mbox{\boldmath$\nabla$}}
\newcommand{\etaperp}{\mbox{\boldmath$\eta_\perp$}}
\newcommand{\bb}{{\bf b}}
\newcommand{\B}{{\bf B}}
\newcommand{\k}{{\bf k}}
\newcommand{\p}{{\bf p}}
\newcommand{\V}{{\cal V}}
\newcommand{\half}{\frac{1}{2}}
\newcommand{\etapar}{{\eta_\parallel}}
\newcommand{\la}{\label}
\newcommand{\I}{\rm Im}
\newcommand{\R}{\rm Re}

\begin{document}

\title{{\bf A derivation of the Boltzmann--Vlasov equation from
multiple scattering using the Wigner function}\thanks
 {This article is dedicated to the memory of Eugene Wigner.}}
\vspace{1 true cm}
\author{J.M. Eisenberg \\ School of Physics and Astronomy \\ 
Raymond and Beverly Sackler Faculty of Exact Sciences \\ 
\vspace {1 true pc}
Tel Aviv University, 69978 Tel Aviv, Israel}
\date{March, 1995}

\maketitle

\begin{abstract}
\baselineskip 1.5 pc
A derivation is given of the Boltzmann--Vlasov equation beginning from
multiple scattering considerations.  The motivation for the
discussion, which is purely pedagogical in nature, is the current
interest in understanding the origins of transport equations in terms of
rigorous field-theory descriptions, or, as in this case, exact 
nonrelativistic formulations.
\end{abstract}
\vfil
PACS: 24.10.Cn, 24.90.+d, 25.70.-z \hfill
\vfil
\newpage
\baselineskip 1.5 pc

\section{Introduction: deriving transport equations \protect \\
from rigorous formalisms}

In 1932 Eugene Wigner introduced \ci{Wig} the concept of what later came 
to be called the Wigner function, which allows a direct and precise 
mapping of quantum dynamics to transport equations and aids the study of
the transition to the classical limit for these 
equations\footnote{Wigner notes \ci{Wig} that the Wigner function
was actually found by Szilard and Wigner some years earlier for a
purpose different from that of the 1932 paper.}.  
This concept has grown up as part of the standard tools of quantum
theory.  In recent years it has become of considerable importance in the
study of relativistic heavy-ion collisions, both for the description of
the dynamics of the
phase in which quarks and gluons are confined in hadrons and for use
in studying the deconfined phase of the quark-gluon plasma.  The general
line of attack in this context
has been to use selected physics ingredients at the level
of an exact formulation in terms of field theory or many-body dynamics
and to exploit the Wigner function in order to derive the
corresponding transport equation.  It is then often possible to
ignore detailed information about phases or fluctuations that is present
in the full quantum theory but is not crucial for an approximate
characterization of the system.  Consequently the transport equation 
serves as a much simpler and much more transparent way of handling the 
basic physics in question.

This general approach raises the question, How well can transport 
methods represent the quantal situation?  One can explore this issue in 
a formal derivation of the transport result beginning from the more
fundamental theory---field theory or many-body theory, say---keeping
track of approximations as one proceeds.  It can also be studied by
comparing numerical simulations of the results of the transport 
formalism with those of the fundamental theory.  Both such methods have
been used in recent years in the study of relativistic heavy-ion physics.
Reviews pertaining to the various possible techniques can be found, for 
example, in refs. [2-6].  

One of the natural techniques that can be used
to generate transport equations for a many-body system---this time a
nonrelativistic one---starts from the multiple-scattering series of
Watson \ci{GW} or of Glauber \ci{Glau} and derives from it the
corresponding transport equation.  This approach has the advantage that
already from the start it is couched in the language of scattering, so
that the progression to a transport formalism is conceptually quite 
natural.  Of course, since the Watson multiple-scattering series can
be derived directly \ci{GW} from the Schr\"odinger equation through 
the Lippmann-Schwinger equation for $N$ fixed potential scatterers,
and since Glauber's series can in turn be derived from Watson's series
\ci{Eis1,Eis2}, such developments of transport theory are intrinsically 
rooted in basic
nonrelativistic quantum theory.  Notably, Thies \ci{Thies} has
systematically studied in full generality the derivation of the Wigner
representation from Watson's multiple-scattering theory.

The purpose of the present brief note is to present a derivation of a
transport equation using Wigner's function and multiple scattering together
with the minimal set of assumptions needed in order to reach the simplest
version of the transport equation, i.e., the analog of the classical
Boltzmann--Vlasov equation.  The motivation for presenting this is 
pedagogical: the complete treatment given by Thies \ci{Thies} 
is quite intricate,
so that there is an advantage to seeing also a more transparent 
development leading specifically to the well-known transport result.
In some ways this makes it easier to focus on the nature of the 
approximations
that are required in order to reach the Boltzmann-Vlasov equation.

\section{Assumptions and formalism}

We address a system in which a particle passes through a medium
of $N$ particles
experiencing optical distortion on the way.  In addition we consider one
direct collision between the projectile and the $i$th particle within the
medium; generalizations to consider $n$ such collisions, 
where $n \ll N,$ are possible.  The target is taken to be a totally 
uncorrelated set of particles,
\be \la{Phi}
|\Phi_0(\r_1,\ldots,\r_N)|^2 = \rho_1(\r_1)\ldots\rho_N(\r_N),
\ee
where the complete lack of correlations implies that the densities for
the individual particles are constants, 
\be
\la{rho}
\rho_i(\r_i) = 1/\Omega,
\ee
where $\Omega$ is
the volume containing the particles.  It is clear that this extreme
assumption  is necessary in order to arrive at the
``standard'' Boltzmann--Vlasov equation at the lowest rung,
where no correlations are entertained, of an anticipated hierarchy.  

We now assume that the projectile possesses sufficiently high energy that
an eikonal form for its wave function is a good approximation,
\bea \la{eikonal}
\phi_\k(\r;\r_i) & = & \exp\left[i\k\cdot\r\right]
\exp\left[-\frac{i}{v}\int_{-\infty}^z\V(\bb,\zeta)
d\zeta\right] \nono \\ 
& \times & \exp\left[-\frac{i}{v}\int_{-\infty}^z 
V_i(\bb-\bb_i,\zeta-z_i) d\zeta\right]. 
\eea
Here $\k$ is the momentum of the projectile, ${\bf v} = \k/m$ is its velocity
with $m$ its mass, $\V$ is the complex potential supplying the optical 
distortion, and $V_i$ is the
interaction between the projectile and the $i$th particle in the medium.
The integrals are taken along the direction of $\k,$ and the position
vectors are taken as $\r = \{\bb,\,z\},$ where $\bb$ is the projection of
$\r$ in a direction perpendicular to $\k$ and $z$ is parallel to $\k.$
Thus the $i$th particle in the target is located at 
$\r_i = \{\bb_i,z_i\}.$
The approximation of extreme high energy in eq. (\ref{eikonal})
is to be expected in order to arrive at a transport equation of classical 
form.  This eikonal form is easily seen \ci{GW,Eis1,EK} to satisfy 
Schr\"odinger's equation approximately at high energy with a potential 
given by the optical potential $\V$ plus the force of the direct 
scattering $V_i.$

Wigner's function \ci{Wig} for this static problem is defined as
\bea \la{Wigner}
W(\r;\,\p) & \equiv & \int d\r_1\cdots d\r_N \, \Phi_0^*(\r_1,\ldots,\r_N)
\int d\bfeta \exp\left[-i\p\cdot\bfeta\right]\phi_\k(\r+\half\bfeta)
\nono \\
& \times & \phi_\k^*(\r-\half\bfeta)\Phi_0(\r_1,\ldots,\r_N) \nono \\
& = & \int d\r_1\cdots d\r_N \, |\Phi_0|^2\ \int d\bfeta 
\ \exp\left[i(\k-\p)\cdot\bfeta\right] \nono \\
& \times &
\exp\left[-\frac{i}{v}\int_{-\infty}^{z+\half\etapar}\V(\bb+\half\etaperp,
\zeta)d\zeta\right]
\exp\left[\frac{i}{v}\int_{-\infty}^{z-\half\etapar}\V^*(\bb-\half\etaperp,
\zeta)d\zeta\right] \nono \\
& \times &
\exp\left[-\frac{i}{v}\int_{-\infty}^{z+\half\etapar-z_i}
V_i(\bb-\bb_i+\half\etaperp, \zeta)d\zeta\right] \nono \\
& \times &
\exp\left[\frac{i}{v}\int_{-\infty}^{z-\half\etapar-z_i}
V_i^*(\bb-\bb_i-\half\etaperp, \zeta)d\zeta\right],
\eea
where $\etapar$ and $\etaperp$ are the components of
$\bfeta$ parallel and perpendicular to $\k.$

In order to generate a transport equation we consider the action of
$v\,\partial/\partial z$ on the Wigner function of eq. (\ref{Wigner}).
We divide the calculation into two parts: first we examine the action
of the derivative on the optical part of the expression, containing $\V,$
to obtain the drift part of the transport equation, and then we let it act
on the direct collision pieces, containing $V_i,$ to
produce the collision term.  The first calculation is very
simple,
\bea \label{drift1}
v\frac{\partial}{\partial z}\, W(\bb,z;\,\p) & = &
-i \int d\r_1\cdots d\r_N \,|\Phi_0|^2\ \int d\bfeta
\ \exp\left[i(\k-\p)\cdot\bfeta\right] \nono \\
& \times &
\left[\V(\bb+\half\etaperp,z+\half\etapar)-
\V^*(\bb-\half\etaperp,z-\half\etapar)\right] \exp[\cdots] \nono \\
& + & {\rm collision\ terms} \nono \\
& \approx & -i \int d\r_1\cdots d\r_N \,|\Phi_0|^2\ \int d\bfeta
\ \exp\left[i(\k-\p)\cdot\bfeta\right] \nono \\
& \times &
\left[2\,i\,\I\,\V(\bb,z)+\bfeta\cdot\bfnabla\,\R\,\V(\bb,z)\right]
\exp[\cdots] + {\rm collision\ terms} \nono \\
& = & \left[2\,\I\,\V(\bb,z) +
[\bfnabla\,\R\,\V(\bb,z)]\cdot\bfnabla_\p\right]\,W(\r,\p) \nono \\
& + & {\rm collision\ terms},
\eea
where we have have indicated only schematically the four exponentials 
of eq.  (\ref{Wigner}) that appear unchanged in each of the subsequent 
expressions.  In eq. (\ref{drift1}) we have also used the usual
approximation that the momentum $\k$ involved here is sufficiently high
so that only small values of $\bfeta$ enter in the integrals.  Once again
this is an approximation that is required for the classical limit to
obtain.  Equation (\ref{drift1}) can be recast in the form
\be \la{drift2}
\left[v\frac{\partial}{\partial z} - 2\,\I\V(\bb,z) 
- [\bfnabla\,\R\V]
\cdot\bfnabla_\p\right]\,W(\r,\p) = {\rm collision\ terms}.
\ee

We now turn to the collision term, omitting explicit reference to the
contributions of the drift term that we have just calculated.  We then
have 
\bea \la{collision1}
v\frac{\partial}{\partial z}\,W(\bb,z;\p) & = &  
-v \int\, d\r_i\, \rho(\r_i)
\ \int d\bfeta\,\exp[i(\k-\p)\cdot\bfeta] \nono \\
& \times &
\frac{\partial}{\partial z_i}\Bigg\{\exp\left[-\frac{i}{v}
\int_{-\infty}^{z+\half\etapar-z_i} 
V_i(\bb-\bb_i+\half\etaperp,\zeta) d\zeta\right] \nono \\
& \times &
\exp\left[\frac{i}{v}
\int_{-\infty}^{z-\half\etapar-z_i} 
V_i(\bb-\bb_i-\half\etaperp,\zeta) d\zeta\right]\Bigg\} \nono \\
& \times & 
\exp\left[-\frac{i}{v}\int_{-\infty}^{z+\half\etapar}\V(\bb+\half\etaperp,
\zeta)d\zeta\right] \nono \\
& \times &
\exp\left[\frac{i}{v}\int_{-\infty}^{z-\half\etapar}\V^*(\bb-\half\etaperp,
\zeta)d\zeta\right],
\eea
where we have replaced the required derivative by $z$ with one by $z_i,$ 
as is easily done in the context of the collision part.  The relevant
limits of integration can be extended from $-\infty$ to $\infty$ if we 
assume that the longitudinal volume dimension
$\Omega_\parallel$ is large relative to the domain of support of the
scattering profiles.
This is then leads to
\bea
\la{collision2}
v\frac{\partial}{\partial z}\,W(\bb,z;\p) & = &
-v \int\, d\bfeta\,\exp[i(\k-\p)\cdot\bfeta]
\int_\Omega\, d^2\bb_i\,dz_i \, \frac{1}{\Omega} \nono \\
& \times &
\frac{\partial}{\partial z_i} \Bigg\{
\exp\left[-\frac{i}{v}
\int_{-\infty}^{z+\half\etapar-z_i} 
V_i(\bb-\bb_i+\half\etaperp,\zeta) d\zeta\right] \nono \\
& \times &
\exp\left[\frac{i}{v}
\int_{-\infty}^{z-\half\etapar-z_i} 
V_i(\bb-\bb_i-\half\etaperp,\zeta) d\zeta\right]\Bigg\}\exp[\cdots]
\eea
where we use eq. (\ref{rho}) explicitly and again suppress the obvious
``inert'' exponentials, this time involving the optical potential.
The integral over $z_i$ is now trivial and we have
\bea
\la{collision3}
v\frac{\partial}{\partial z}\,W(\bb,z;\p) & = & 
-v \int\, d\bfeta\,\exp[i(\k-\p)\cdot\bfeta]
\int_{\Omega_\perp}\, d^2\bb_i \, \frac{1}{\Omega} \nono \\
& \times &
\Bigg\{1 - \exp\left[-\frac{i}{v} \int_{-\infty}^\infty 
V_i(\bb-\bb_i+\half\etaperp,\zeta) d\zeta\right] \nono \\
& \times &
\exp\left[\frac{i}{v} \int_{-\infty}^\infty 
V_i(\bb-\bb_i-\half\etaperp,\zeta) d\zeta\right]
\Bigg\}\exp[\cdots] \nono \\
& = & v \int\, d^2\bfeta_\perp\ 
\int_{-\infty}^\infty\,d\eta_\parallel\ 
\int\frac{d\p'}{(2\pi)^3}\ 
\exp\left[i({\p'}_\perp-\p_\perp)\cdot\etaperp\right] \nono \\
& \times &
\exp\left[i({p'}_\parallel-p_\parallel)\etapar\right]
\int_{\Omega_\perp} d^2\bb_i\, \frac{1}{\Omega} \nono \\
& \times &
\Bigg\{[1 + \Gamma_i(\bb-\bb_i+\half\etaperp)] 
[1 + \Gamma_i^*(\bb-\bb_i-\half\etaperp)] - 1 \Bigg\} \nono \\
& \times & W(\r;\,\p'),
\eea
where the integration over the transverse variable $\bb_i$ ranges
over the transverse cross section $\Omega_\perp$ of the volume 
$\Omega,$  and we have defined the usual profile functions \ci{Glau,EK},
\be
\la{Gamma}
\Gamma_i(\B) \equiv 
\exp\left[-\frac{i}{v}\int_{-\infty}^\infty\,
V_i(\B,\zeta)d\zeta\right] - 1.
\ee
Then
\bea
\la{collision4}
v\frac{\partial}{\partial z}\,W(\bb,z\,;\p) & = &
2\pi v\,\int\frac{d\p'}{(2\pi)^3}\ 
\delta(p'_\parallel - p_\parallel)\ \int\,d^2\etaperp\ 
\int_{\Omega_\perp}\,d^2\bb_i\ \, \frac{1}{\Omega} \nono \\
& \times &
\Bigg\{[1 + \Gamma_i(\bb-\bb_i+\half\etaperp)]
[1 + \Gamma_i^*(\bb-\bb_i-\half\etaperp)] - 1 \Bigg\} \nono \\
& \times &
\exp\left[i(\p'_\perp - \p_\perp)\cdot\etaperp\right]\ W(\r;\,\p')
\nono \\
& = & 2\pi v\,\int\frac{d\p'}{(2\pi)^3}\ 
\delta(p'_\parallel - p_\parallel)\,\int\,d^2\B\,d^2\B' \nono \\ 
& \times &
\,\frac{1}{\Omega} 
\bigg\{[1 + \Gamma_i(\B)][1 + \Gamma_i^*(\B')] - 1 \bigg\} \nono \\
& \times &
\exp\left[i(\p'_\perp-\p_\perp)\cdot(\B-\B')\right]\ W(\r;\,\p').
\eea
The two-dimensional Fourier-transform integral over $\B$ and $\B'$ 
containing both profile functions immediately
yields \ci{Glau,EK} the scattering amplitudes at momentum 
transfer $\p'_\perp-\p_\perp,$ assuming the transverse volume
$\Omega_\perp$ is large relative to the domain of support of the
scattering profile.  Then
\bea
\la{collision5}
v\frac{\partial}{\partial z}\,W(\bb,z;\p) & = &
2\pi v\,\int\frac{d\p'}{(2\pi)^3}\ 
\delta(p'_\parallel - p_\parallel) \left(\frac{2\pi}{k}\right)^2\ 
\frac{d\sigma_i(\p'\to\p)}{d\Omega} \nono \\
& \times &
W(\r;\,\p')\,\rho_i(\r),
\eea
where we have suppressed terms with transverse delta functions 
which do not contribute.  We finally arrive at
\be
\la{collision6}
v\frac{\partial}{\partial z}\,W(\bb,z;\,\p) = 
\frac{1}{mk}\,\int\,d^2\p'_\perp\ 
\frac{d\sigma_i(\p'\to\p)}{d\Omega}\ W(\r;\,\p'_\perp,p_\parallel)\ 
\rho_i(\r)
\ee
for the collision term.

Putting the drift and collision pieces together we arrive at the 
Boltzmann-Vlasov transport equation
\bea
\la{BV}
\bigg[v\frac{\partial}{\partial z} - 
2\,\I\V & - & \frac{\partial\,{\R}\V}{\partial z}\,
\frac{\partial}{\partial p_\parallel}\bigg] 
W(\bb,z;\,\p'_\perp,p_\parallel) = \nono \\
& = & \int\ \frac{d^2\p'_\perp}{mk}\ 
\frac{d\sigma_i(\p'\to\p)}{d\Omega}\ 
W(\bb,z;\, \p'_\perp,p_\parallel) \rho_i(\r),
\eea
which is the result arrived at by Thies \ci{Thies} as the lowest-order
limit of a complete, but necessarily much more intricate,
development.\footnote{We note that with the present method it is 
unnecessary to assume \ci{Huef} that the profile function always
leads to forward contributions in the sense that
$\Gamma(\bb-\bb_i,\,z-z_i) = \Gamma(\bb-\bb_i)\,\Theta(z-z_i).$}

\section{Conclusions}

Our result is wholly contained in eq. (\ref{BV}), and is the 
entirely expected Boltzmann--Vlasov equation with the role of the 
drift potential played by the background optical potential and 
explicit scattering appearing on the right-hand side as a collision
term.  This result is derived here in a relatively simple manner
since we have been prepared to restrict ourselves from the very
start to lowest-order features.  In order to produce this
classical transport equation our main assumptions have been 
(i) that the projectile enters the system at extremely high energy;
(ii) that the scatterers within the bombarded system are completely
without correlation; and (iii) that they are contained in a volume
much larger than the support domain of the individual scatterings.

\vskip 2 true pc

It is a pleasure to acknowledge conversations with D.S. Koltun which
led to my initial interest in transport equations in medium-energy
physics.  This work was funded in part by the U.S.-Israel Binational
Science Foundation and in part by the Ne'eman Chair in Theoretical 
Nuclear Physics at Tel Aviv University.

\vskip 2 true pc

\end{document}